\documentclass[preprint,superscriptaddress,preprintnumbers,a4paper,amsmath,amssymb,showpacs,floatfix]{revtex4}
\usepackage{graphics}
\usepackage{graphicx}
\usepackage{bm}

\begin{document}

\title{High temperature tetragonal crystal structure of UPt$_2$Si$_2$}

\author{K.~Proke\v{s}}
\email{prokes@helmholtz-berlin.de}
\affiliation{Helmholtz-Zentrum
Berlin f\"{u}r Materialien und Energie, Hahn-Meitner Platz 1, 14109 Berlin, 
Germany}

 \author{O. Fabelo}
  \affiliation{Inst. Laue Langevin, 71 Ave Martyrs,CS 20156, F-38042 Grenoble 9, France}
   
 \author{S. S\"{u}llow}
 \affiliation{Institut f\"{u}r Kondensierte Materie, TU Braunschweig, 38106 Braunschweig, Germany}
 
 \author{J. Lee} 
 \affiliation{CALDES, Institute for Basic Science, Pohang 37673, Republic of Korea}
  
 \author{J.A.~Mydosh}
 \affiliation{Institute Lorentz and Kamerlingh Onnes Laboratory, Leiden University, Leiden 2300 RA , The Netherlands}
  
\date{\today}

\begin{abstract}
High temperature crystal structure of UPt$_2$Si$_2$ determined using single-crystal neutron diffraction at 400 K is reported. It is found that the crystal structure remains of the primitive tetragonal CaBe$_2$Ge$_2$ type with the space group $P$4/$n m m$. Anisotropic displacement factors of the Pt atoms at the 2$a$ (3/4 1/4 0) and Si atoms at the 2$c$ (1/4 1/4 $z$) Wyckoff sites are found to be anomalously large.

\end{abstract}

\maketitle

\section{Introduction} 

Uranium based intermetallic compounds have been a focus of an intensive research for many years due to their unique physical properties that arise mainly from a large extent of unfilled 5$f$-electron wave functions. \cite{Sech98} Their vicinity to the Fermi level and spatial extent leads to a sensitivity of these materials to external perturbations and existence of a vast number of crystallographic modifications. Many of the magnetic and bonding phenomena are due to a hybridization between 5$f$ states and other electronic states in the solid and which controls for instance whether the system has localized or itinerant 5$f$ magnetic states. As the hybridization depends on the number, distance of neighboring atoms (called ligands) and the geometry surrounding the 5$f$ element it is extremely important to determine the details of the crystal structure. \cite{Sech98,Hoffmann85} Tetragonal UT$_2$X$_2$~compounds (T stands for a late a transition metal and M for Si or Ge) comprising more than 20 members has been in the focus of scientific community for more than the last three decades. The most prominent member of the group, URu$_2$Si$_2$ exhibits an unconventional superconductivity \cite{Palstra85,Hasselbach91,Mydosh14} coexisting together with another, yet unknown, type of order \cite{Mydosh14}. Most of these compounds adopt the ThCr$_2$Si$_2$~type of structure (space group $I$4$m m m$). \cite{Sech98,Hoffmann85} However, few of them, namely those with T = Ir or Pt are reported to crystallize in the $klassengleiche$ centrosymmetric subgroup CaBe$_2$Ge$_2$ type of structure with space group $P$4/$n m m$. \cite{Sech98,Bak85} Common for both these structures is the absence of ($h$ 0 0): $h$ = 2$n$ + 1 and ($h$ $k$ 0 ): $h$ + $k$ = 2$n$ + 1 reflections that are systematically extinct due to an $n$-type glide plane.

The subject of this work is the confirmation of the UPt$_2$Si$_2$ crystal structure that is reported to adopt the latter, primitive version over the entire temperature range. \cite{Bak85,Steeman88,Bleckmann10} UPt$_2$Si$_2$ has been previously subject of numerous studies concentrated on the magnetic, transport and thermal properties. Particular attention has been devoted to the crystal and magnetic structures. \cite{Bak85,Steeman88,Bleckmann10,Amitsuka92,Sullow08,Grachtrup17,Otop04} UPt$_2$Si$_2$ orders antiferromagnetically (AF) below T$_{N}$ that is reported to range between 32 K and 35 K. \cite{Sullow08,Steeman88} It has been reported that the low-temperature AF structure is collinear with U moments of 1.9 - 2.5 $\mu_{B}$, depending on a study,\cite{Sullow08,Steeman88} pointing along the $c$-axis direction.  It can be described as a (up-down) stacking of ferromagnetic uranium sheets alternating along the tetragonal axis direction. Upon application of high magnetic fields it exhibits field-induced transitions for both, the $a$ and $c$-axis directions with critical fields of about 40 T and 24 T, respectively.\cite{Sech98,Bak85,Steeman88,Bleckmann10,Amitsuka92,Sullow08} The field transition has been interpreted as a topological Lifshitz transition.\cite{Grachtrup17} Much attention has been paid to a possible crystallographic disorder \cite{Sullow08} that has been linked to irreversibility in magnetic behavior.\cite{Otop04} Neutron powder and single crystal diffraction experiments have indicated that the one of the two Pt and Si atomic positions exhibit anomalously large displacement parameters.\cite{Sullow08} 

\begin{figure}
\includegraphics*[scale=0.3]{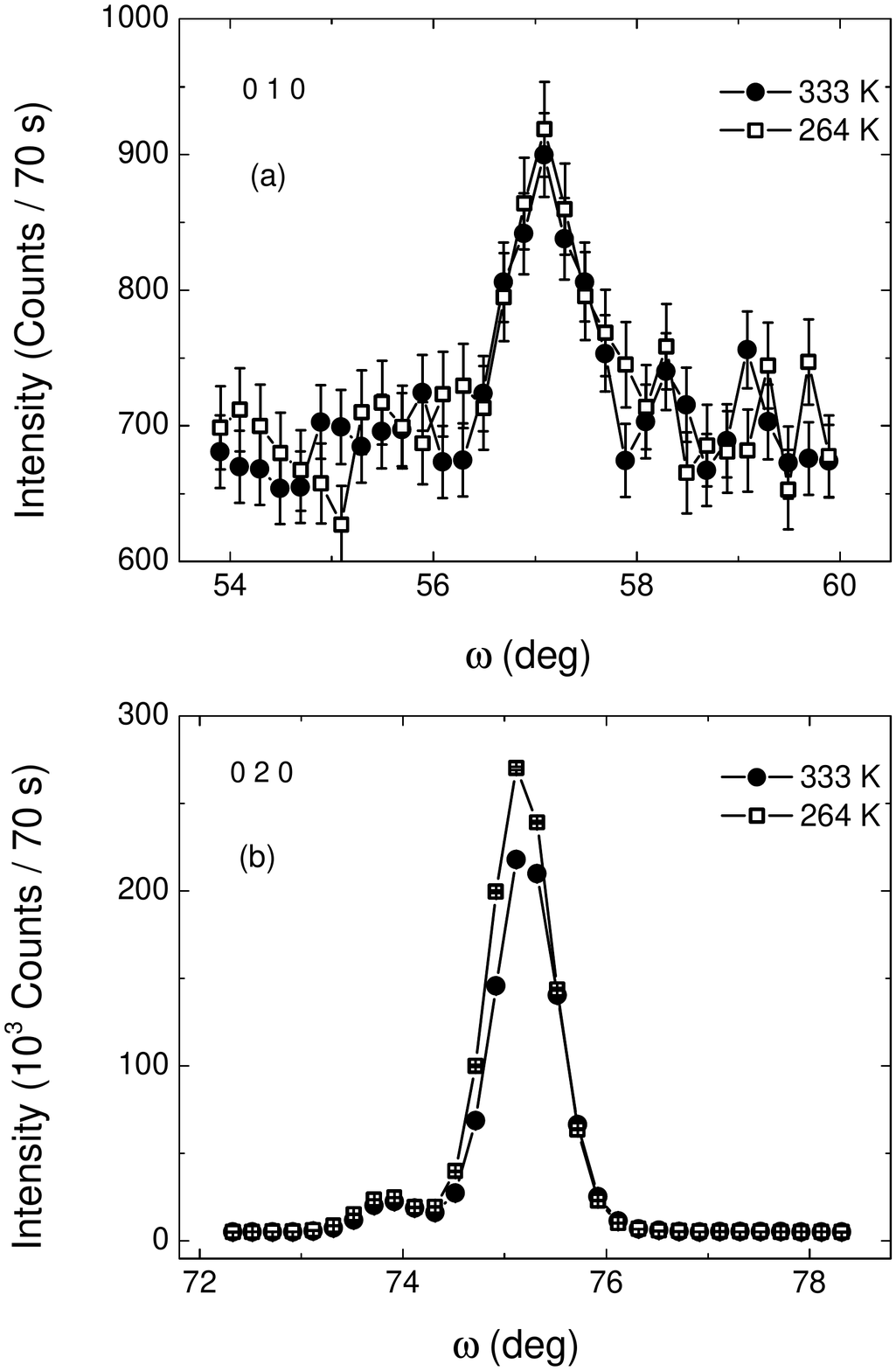}
\caption{Rocking curves through (a) the 010 and (b) the 020  reciprocal position of the UPt$_2$Si$_2$ single crystal recorded above T$_{s}$ $\approx$ 315 K, at 333 K (closed points) and below T$_{s}$ at 264 K (open points), respectively.} 
\label{fig1}
\end {figure}

UPt$_2$Si$_2$ exhibits near the room temperature, around T$_{s}$ $\approx$ 315 K, a significant specific heat anomaly. A discontinuity at similar temperature is seen also in the temperature dependence of the electrical resistivity. \cite{Bleckmann10} As powder diffraction experiments did not revealed any significant changes in diffraction patterns, the origin of these anomalies remained for a long time a mystery. We then speculated that it is due to a crystal structure modification appearing at  T$_{s}$. However, recently we have detected above T$_{s}$ using a large UPt$_2$Si$_2$  single crystal glued to a sample holder very weak Bragg reflections violating the ($h$ 0 0): $h$ = 2$n$ + 1 (see Fig.~\ref{fig1} (a)) and ($h$ $k$ 0 ): $h$ + $k$ = 2$n$ + 1 extinction rules. The typical intensity ratio between these forbidden and allowed reflections was 1 : 1000, i.e. much larger than possible contamination due to $\lambda$/2 contamination of the E4 neutron instrument. We continued speculating that the space group given in the literature is in error and that the glide plane connecting two of the Pt (and two Si) atoms is missing lowering the space group to $P$ -4 $2~m$ thereby splitting these sites into two independent sites. Different models were proposed, among them a different degree of occupation, mixing of atoms, different thermal factors at the two split sites and/or a subtle orthorhombic distortion have been considered. All these mechanisms would lead to an existence of forbidden reflections. The proximity of the structural transition leads to yet another possible scenario that the observed forbidden reflections are due to strain in the sample caused by the stiff glue. To be able to distinguish between different models we have undertaken a diffraction experiment on a relatively fresh small single crystal that was not cycled through the structural transition many times. Nevertheless, in order to determine the crystal structure at elevated temperatures, above the structural transition required a collection of a reasonable number of Bragg reflections, such a task is possible using the 4-circle instrument D9 at Institute Laue-Langevin (ILL), Grenoble. Neutron diffraction (instead of X-ray) has been used due to the presence of a heavy uranium that would make the x-ray determination difficult. Herein we report single crystal results of such a neutron diffraction study, which unambiguously confirms that UPt$_2$Si$_2$  forms in the centrosymmetric CaBe$_2$Ge$_2$ type structure.

\section{Experimental} 

Single crystal of UPt$_2$Si$_2$ has been prepared by a modified tri-arc Czochralski method at the Leiden University. Characterization, including specific heat and high-field measurements can be found in literature. \cite{Menovsky86,Sullow08,Bleckmann10} For the diffraction experiments we have used two pieces originating from the same original large crystal. While at the Helmholtz-Zentrum Berlin (HZB) we have used $\approx$ 1.5 g heavy crystal of an irregular shape; at ILL a small bar-shaped, about 21 mg heavy crystal with dimensions 1 x 1 x 2 mm$^{3}$ with the longer dimension along the $c$-axis direction cut using spark-erosion has been used. Two separate diffraction experiments were carried out. At HZB we have used double-axis diffractometer E4 that that is equipped with a 200 x 200 mm$^3$ position sensitive detector (PSD). The incident neutron wavelength of 2.41 \AA~produced by the Pyrolythic graphite PG(002) monochromator has been used. \cite{Prokes17} Here we have collected data (typically 30 s per point) at temperatures from $\approx$ 260 K, i.e., below the specific heat anomaly up to $\approx$ 330 K, well above the T$_{s}$. At the ILL, the D9 hot neutron diffractometer with the incident wavelength $\lambda$ = 0.8377 \AA, operated in the four-circle mode has been utilized. While in the former experiment the crystal has been glued to an aluminium holder, for the ILL experiment the crystal has been wrapped in an aluminium foil that has been fixed on the sample holder pin using dedicated high-temperature glue. In this way we have minimalized in the latter case the stress acting on the sample. Data were at D9 collected well above the structural transition temperature, at 400 K, achieved using a four-circle furnace. In order to be able to determine the crystal structure of UPt$_2$Si$_2$, we have recorded at D9 in total, 457 Bragg reflections, 292 of them unique. Among those were also 22 Bragg reflections that are forbidden for the space group $P$4/$n m m$. Allowed reflections were recorded with time 5 sec per point, forbidden ones mostly with 128 sec per point, in part with 300 sec per point. At both instruments, in order to reduce the $\lambda$/2 contamination, we have used additional filters. In both experiment, the orientational UB matrix has been calculated from several well centered Bragg reflections.

The crystal structure refinement has been performed using computer codes Fullprof \cite{Fullprof01}. Standard nuclear scattering lengths $b$(Pt) = 9.60 fm, $b$(Si) = 4.149 fm and $b$(U) = 8.417 fm were used. \cite{Sears92} For the absorption correction (Gaussian integration) we used the absorption coefficient $\mu$ = 0.15 cm$^{-1}$. Anisotropic empirical extinction correction as implemented in Fullprof has been applied. Crystal structure has been drawn using graphical computer code VESTA. \cite{VESTA11} 

\section{Neutron Diffraction Results} 

In Fig.~\ref{fig1} (a) we show rocking curves through the 010 reciprocal position of the UPt$_2$Si$_2$ single crystal recorded above T$_{s}$ at 333 K (closed points) and at 264 K (open points), respectively. As can be seen, at both temperatures we observe the 010 Bragg reflection. This observation is rather striking as it suggests that the published crystal structure of UPt$_2$Si$_2$ is not correct. Namely, the ($h$ 0 0): $h$ = 2$n$ + 1 type reflections are forbidden for the $P$4/$n m m$ space group. 

\begin{figure}
\includegraphics*[scale=0.4]{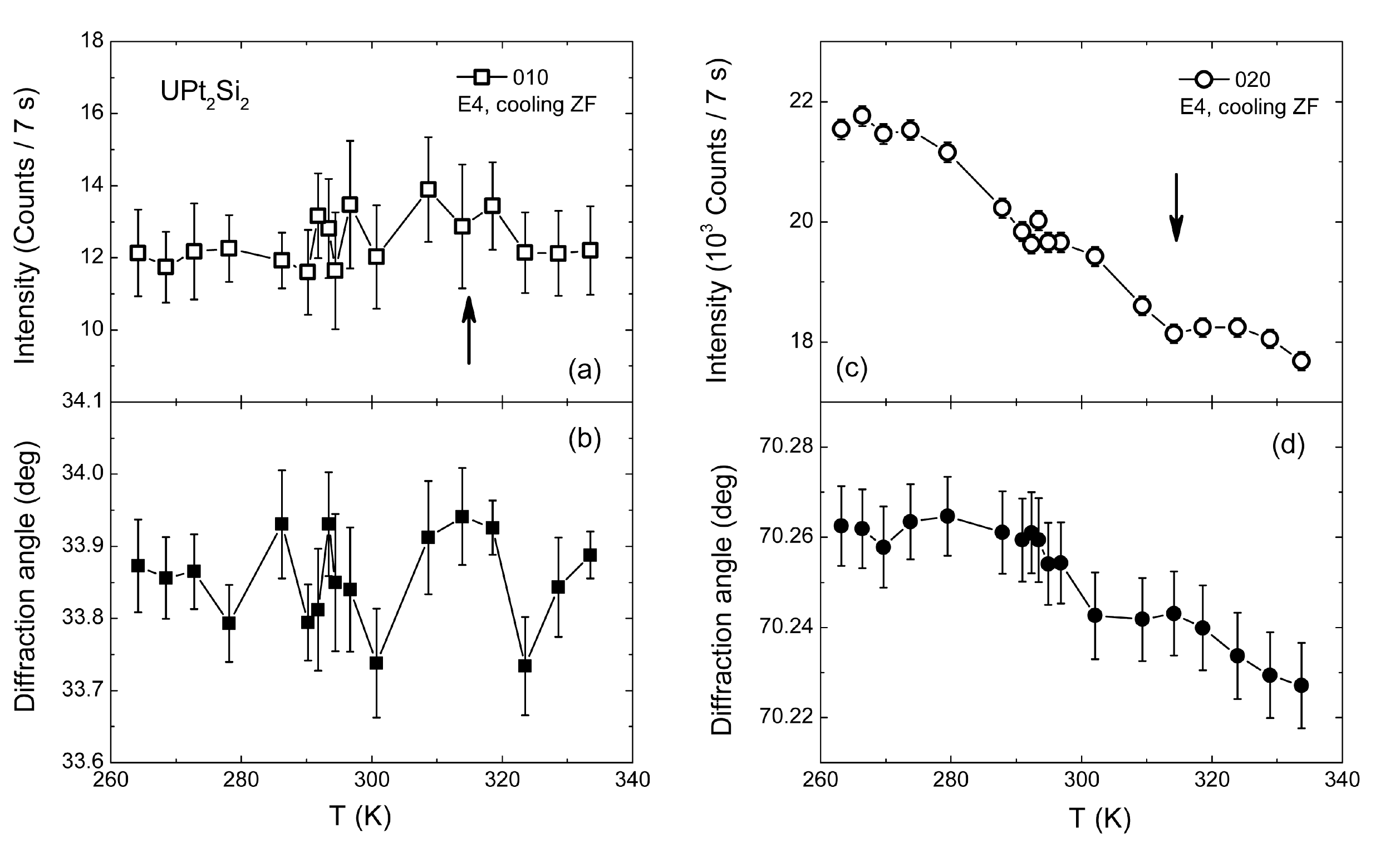}
\caption{Temperature dependence of the integrated intensities of (a) the 010 and (c) the 020 Bragg reflections measured on the large UPt$_2$Si$_2$~single crystal at E4 upon cooling across the anticipated crystal structure transition along with the respective diffraction angle at which the reflections appear ((b) and (d), respectively). Arrows denote the position of the anomaly seen on the temperature dependence of the specific heat that takes place around 315 K. } 
\label{fig2}
\end{figure}

While the intensity of the 010 Bragg reflection hardly changes across the T$_{s}$, Fig.~\ref{fig1} (b) documents that the 020 reflection increases. Indeed, as shown in Fig.~\ref{fig2} (a) and (c), the temperature dependences of the integrated intensities of these two reflections suggest that the 010 is temperature invariant whereas the 020 increases upon cooling. The different behavior of the two reflections is also seen in their diffraction angle value as shown in Fig.~\ref{fig2} (b) and (d). While the former reflection hardly changes its position, the 020 Bragg reflection shifts upon cooling to higher diffraction angles suggesting shrinking of the $a$-axis lattice parameter. Different behavior of the two Bragg reflections suggest that the 010 is not due to a finite $\lambda$/2 contamination although a non-zero contamination cannot be excluded even despite use of an additional filters. In addition, the increase of the 020 integrated intensity upon cooling from 333 K to 264 K is too large to be explained by the reduction of the displacement parameters.  We propose that it is due to a crystal structure modification upon cooling across $T_{s}$. However, details of this transition will be published elsewhere. \cite{Lee20}
     
In order to verify the existence/non-existence of the ($h$ 0 0): $h$ = 2$n$ + 1 type reflections at elevated temperatures, we have performed the experiment using the D9 instrument. During this experiment we have paid attention to the ($h$ 0 0): $h$ = 2$n$ + 1 type reflections. In few cases a small intensity has been detected at these, for the $P$4/$n m m$ space group forbidden reflection positions. These, however, have been proven to be due to small residual $\lambda$/2 contamination. These reflections vanished when an additional erbium $\lambda$/2 filter has been used.

    \begin{figure}
\includegraphics*[scale=0.25]{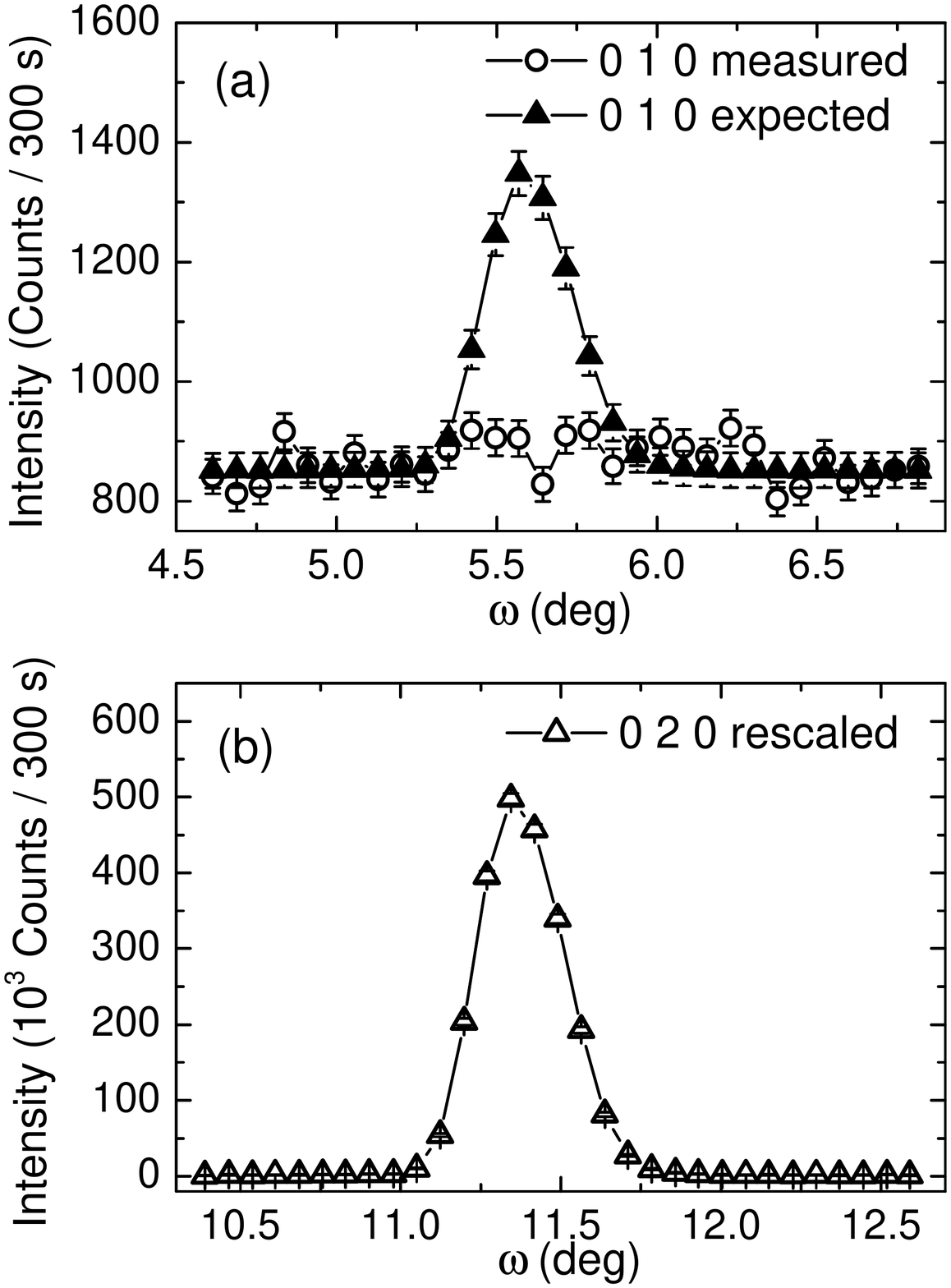}
\caption{(a) Rocking curve through the 0 1 0 reciprocal position of the UPt$_2$Si$_2$ single crystal recorded at 400 K using D9 instrument at ILL with measurement time of 300 s per point (open symbols) together with re-scaled intensity of the 0 2 0 Bragg reflection divided by a factor of 1000 (closed symbols), i.e. by the  intensity factor between these reflections as observed in E4 experiment. (b) Experimental rocking curve through the 0 2 0 reciprocal position rescaled for the same time as in the (a) panel. } 
\label{fig3}
\end 
{figure}

In Fig.~\ref{fig3} (a) we show the experimental rocking curve through the 010 reciprocal point as recorded at D9 instrument at 400 K with measurement time of 300 sec per point utilizing an additional Er filter together with the expected intensity calculated from the rocking curve through the 020 Bragg reflection (see  Fig.~\ref{fig3} (b)). The latter, expected, signal has been calculated with an assumption that the ratio between the 010 and 020 reflections would be similar as in the case of the E4 experiment. Clearly, we do not observe any intensity at the 010 reciprocal space position. This result is in disagreement with the experimental evidence from the E4 experiment.

\section{Structure refinement}

In the course of the refinement, we have tested several crystal structure models. In particular, we have tried to to test solutions that allow for the ($h 0 0 $): $h$ = 2$n$ + 1 and ($h k 0 $): $h + k$ = 2$n$ + 1 type of reflections. Although not observed in the D9 experiment, E4 experiment has revealed the existence of these reflections. One has to survey all the non-isomorphic subgroups that do not contain the glide plane leading to extinction of above mentioned forbidden reflections. There are only two such maximal subgroups, namely the $P 4 m m$ and $P \bar{4} m 2$. We have looked for solutions that are close to the original  CaBe$_2$Ge$_2$~type of structure. While we could not find any reasonable model with the former space group, with the latter it is possible to construct exactly this crystal structure. Within this space group splits original Pt1 and Si1 atomic positions into two inequivalent sites. However, although the extinction rules for the $P \bar{4} m 2$ space group allow for all the ($h 0 0 $) and ($h k 0 $) reflections, these reflections would be still extinct if the atoms would occupy these sites with the same occupation and the same thermal displacement factors. In order to develop non-zero intensities on these reflections, these sites would need to be different. There are few possibilities to achieve this. First, the occupation could be different in terms of the percentage of atoms taking the respective sites. Second, there might be a finite preferential occupation or mixing of atoms among different sites. Third, the thermal displacement factors, isotropic or anisotropic, at different sites could be different. Combinations of the effects are, of course, possible as well. Nevertheless, it appears that the quality of fits allowing for these different possibilities is not better than conserving the original CaBe$_2$Ge$_2$~type of structure. Also attempts assuming a small orthorhombic distortion appear not to be significantly better. 

\begin{figure}
\includegraphics*[scale=0.25]{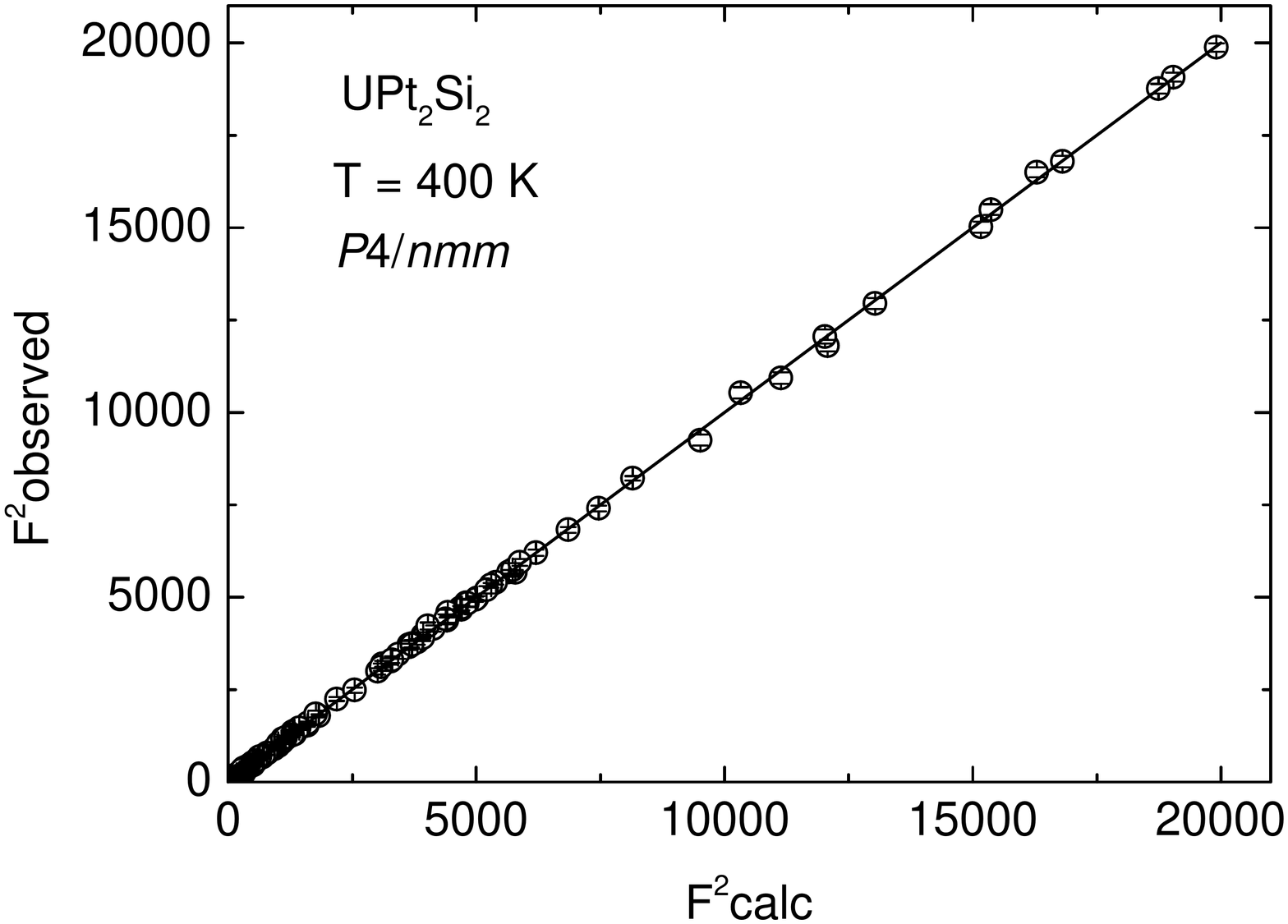}
\caption{Calculated versus observed structure factors squared recorded on UPt$_2$Si$_2$  single crystal at 400 K using D9 instrument at ILL.} 
\label{fig4}
\end 
{figure}

\begin{figure}
\includegraphics*[scale=0.25]{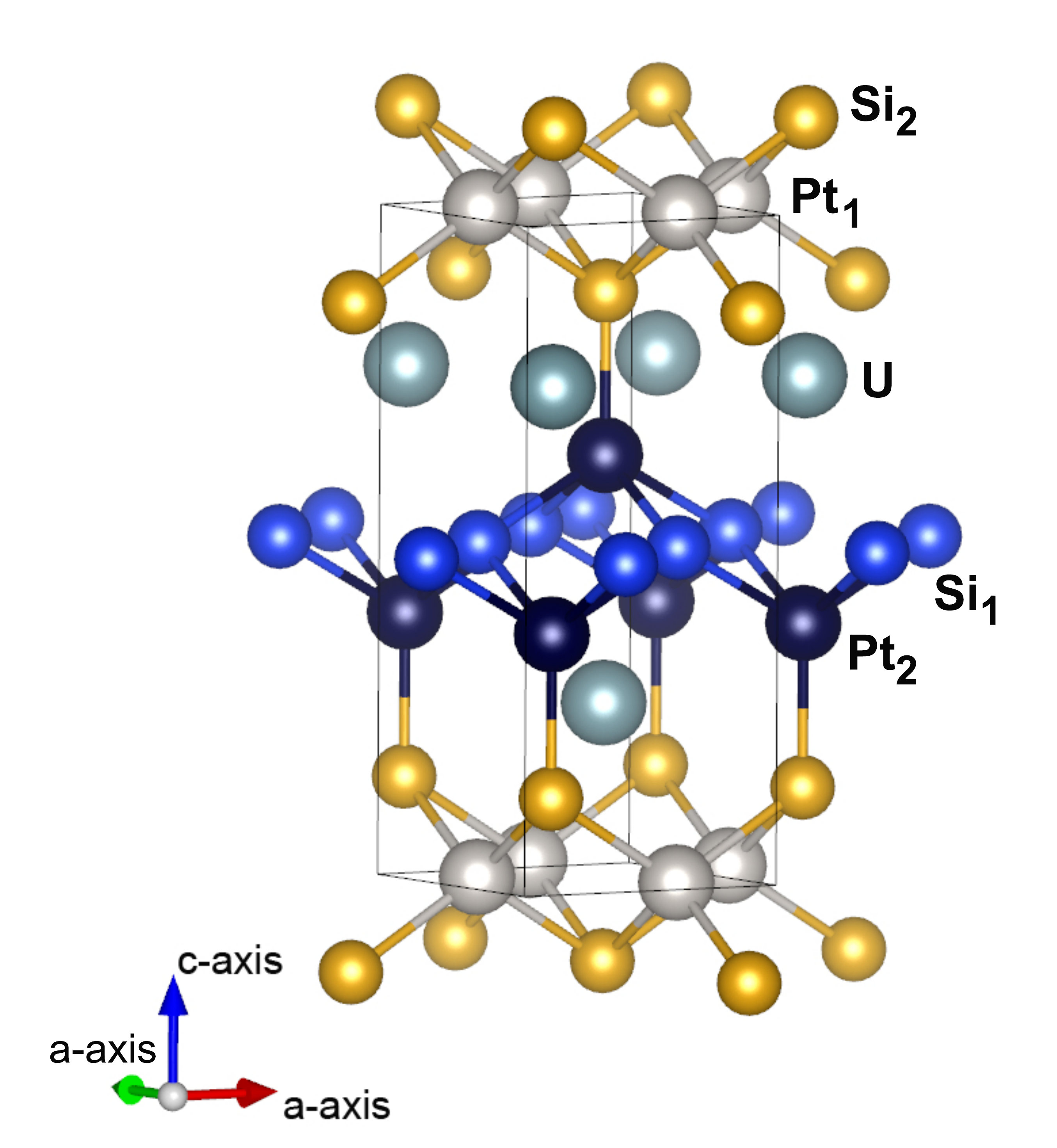}
\caption{A schematic representation of the tetragonal crystal structure of UPt$_2$Si$_2$ (CaBe$_2$Ge$_2$ type, space group $P$4/$n m m$) as determined from neutron diffraction data taken at D9 at 400 K. U atoms are shown as large (light blue), Pt atoms as intermediate (gray and dark blue), and Si atoms as small (blue and orange) spheres, respectively. Bonds highlight the nearest neighbor Pt-Pt, Si-Si and Pt-Si distances.} 
\label{fig5}
\end{figure}

The collected data set \cite{ILLDOI} described in Tab. \ref{tab:table1} thus appear to be in agreement with the primitive tetragonal lattice. In accordance with the observed systematic extinctions, the centrosymmetric space group $P$4/$n m m$ was found to be correct during the structure refinement. The starting positional parameters were obtained the literature. \cite{Bak85,Steeman88,Bleckmann10} Subsequently, the structural parameters were refined on the squared structure factors $F^{2}$ using anisotropic displacement parameters (ADPs) from all atoms. Plot of the observed versus calculated nuclear structure factors of UPt$_2$Si$_2$~after the extinction correction is shown in Fig.~\ref{fig4}. The schematic representation of the UPt$_2$Si$_2$~crystal structure is shown in in Fig.~\ref{fig5}. The structure refinement reveals that the structure of UPt$_2$Si$_2$ is of the CaBe$_2$Ge$_2$ type. 

\begin{table}
\caption{\label{tab:table1}Crystallographic data of UPt$_2$Si$_2$~single crystal and structure refinement information.}
\begin{tabular}{ll}
Empirical formula & UPt$_2$Si$_2$ \\
Molar weight (g mol$^{-1}$) & 684.36  \\
Crystal system & Tetragonal \\
Space group, Z & $P$4/$n m m$, 2 \\
Temperature (K) & 400 \\
Unit cell dimensions (\AA) & $a$ = 4.2056(7) \\
& $c$ = 9.709(4) \\
Calculated density (g cm$^{-3}$)& 13.23\\
Crystal size (mm$^{3}$)& 1$\times$1$\times$2 \\
Absorption coefficient (cm$^{-1}$) & 0.15 \\
$\Theta$ range ($\deg$) & 9-85 \\
Range in $hkl$ & 0+6, 0+4, -2+10\\
Number of reflections measured & 457\\
Independent reflections measured & 292\\
Observed reflections & 216 \\
Number of reflections with I > 2$\sigma$& 185\\
Number of free parameters & 20\\
Number of restrains & 0\\
Goodness of fit $F^{2}$ & 1.33\\
$RF^{2}w$-factor& 2.20\\
$\chi^{2}$ & 0.89 \\
\hline
  \end{tabular}
\end{table}

The best fit to the data (see Fig.~\ref{fig4}) leads the crystallographic parameters of the UPt$_2$Si$_2$ crystal structure, in particular the positional and the temperature displacement parameters, listed in Tab. \ref{tab:table2}. The crystal structure is schematically shown in Fig.~\ref{fig5}. The agreement between the observed and calculated structure factors is excellent, leading $R_{F}$ = 2.78 if only isotropic displacement parameters factors and $R_{F}$ = 1.86 if ADP are considered, respectively. Negligible occupational defficiency of the order of 1 \% has been fount at the Pt1 2$a$ (3/4 1/4 0) site. 

\begin{table}
\caption{\label{tab:table2}Crystal structure parameters of a UPt$_2$Si$_2$~single crystal determined from neutron diffraction data set obtained above T$_{s}$, at 400 K at D9 (ILL). For symmetry reasons are $U_{11}$ = $U_{22}$ and the values $U_{12}$, $U_{13}$, and $U_{23}$ of all the atoms in the structure are equal to zero. $U_{eq}$ is defined as one third of the trace of the orthogonalized $U_{ij}$. Standard deviations are given in parentheses.}

\begin{tabular}{lcccccccc}
  \hline
  \hline
\multicolumn{3}{l}{$T$ = 400 K} & & & \multicolumn{2}{l}{S.G. : $P$4/$n m m$}\\
Atom& Site & $x/a$ & $y/a$& $z/c$ & occ. &$U_{11}$ & $U_{33}$ & $U_{eq}$\\
& &  & &  & & (\AA$^{-2}$) & (\AA$^{-2}$) & (\AA$^{-2}$) \\
\hline
  U& 2$c$ &1/4 & 1/4 & 0.74833(4) & 1.00(f) & 0.0073(2) &  0.0014(1) & 0.0066(2) \\
  Pt1&2$a$&3/4 & 1/4 & 0 &   0.989(3) & 0.0175(2) &  0.0021(1) & 0.0139(2)  \\
  Pt2&1$c$&1/4 & 1/4 &  0.37868(4) &  1.001(1) & 0.0067(2) &  0.0015(1) & 0.0063(2)  \\
  Si1&2$b$&3/4 & 1/4 & 1/2 &  1.000(1) &0.0076(3) &  0.0019(1) & 0.0076(2)  \\  
  Si2&2$c$&1/4 & 1/4 & 0.13357(4) &  0.996(2) & 0.0123(4) &  0.0014(2) & 0.0095(3)  \\
 \hline
\hline
\multicolumn{3}{l}{Agreement factor:} & $R_{F}$ = & 1.86 \\
\hline

  \end{tabular}
\end{table}

The positional parameters refined given in Tab. \ref{tab:table2} are very close to the literature values. \cite{Bleckmann10} Inspection of the refinement parameters indicates an anomalous enhancement of the ADP for Pt1 atoms at the 2$a$ and Si2 atoms at the 2$c$ positions. These are, however, somewhat smaller than in the literature. \cite{Bleckmann10} 
In Fig. ~\ref{fig5} we also show the shortest Pt-Pt, Si-Si and Pt-Si bonds. The two crystallographically independent Pt atoms are coordinated differently by Si atoms that occupy also two independent sites. The Pt1 atoms are coordinated tetrahedrally by Si2 atoms, whereas the five neighbors of Pt2 are arranged in a square pyramid intersecting the plane made of U atoms. Interatomic distances are listed in Tab. \ref{tab:table3}. As can be seen, the Pt–Si distances range from 238 to 247 pm, i.e. somewhat smaller than the sum of the covalent radii for Pt + Si of 247 pm. \cite{Emsley99} This illustrates to a very strong bonding between Pt and Si. The largest neighbor Pt-Si distance is found between for Pt1 atoms at the 2$a$ and Si2 atoms at the 2$c$ positions, for which also an enhanced ADP values are found (see Tab. \ref{tab:table2}).  Let us note that the same atoms (Pt1 and Si2) are constitute a buckled layer (see Fig. ~\ref{fig5}) suggesting that these pairs of atoms might be susceptible to severe distance modifications thus making the crystal structure potentially unstable. Nevertheless, our present data do not indicate any form of modification/distortion from the CaBe$_2$Ge$_2$ type structure.

\begin{table}
\caption{\label{tab:table3}Interatomic distances in UPt$_2$Si$_2$ as determined from the D9 neutron diffraction experiment at 400 K.}
\begin{tabular}{lllllllll}
Atom & Neighbor & Number & Distance & &Atom & Neighbor & Number & Distance \\
 &  &  & (\AA) & &  &  &  & (\AA)  \\
  \hline
  U: & Si2 & 4 & 3.187 & & Pt2: & Si2 & 1 & 2.380 \\
 & Si1 & 4 & 3.199 & & & Si1 & 4 & 2.410 \\
 & Pt2 & 4 & 3.219 & & & U & 1 & 3.219 \\
 & Pt1 & 4 & 3.224 & & & U & 4 & 3.589 \\
 & Pt2 & 1 & 3.589 & & Si1: & Pt2 & 4 & 2.410 \\
 & Si2 & 1 & 3.740 & & & Si1 & 4 & 2.974\\
 & U & 4 & 4.206 & & & U & 4 & 3.199\\
Pt1: & Si2 & 4 & 2.471 & & Si2: & Pt2 & 1 & 2.380 \\
  & Pt1 & 4 & 2.974& & & Pt1 & 4 & 2.471\\
 & U & 4 & 3.224 & & & U & 4 & 3.187\\
 &  &  &  & & & U & 1 & 3.740 \\
  \hline
  \end{tabular}
\end{table}

\section{Conclusions} 

It is now clear that UPt$_2$Si$_2$ adopts at high temperature above the structural transition around $T_{s}$ the CaBe$_2$Ge$_2$ type structure with the space group is $P$4/$n m m$. This result is in agreement with the literature \cite{Sech98,Hoffmann85}. The existence of the forbidden reflections of the ($h$ 0 0): $h$ = 2$n$ + 1 and ($h$ $k$ 0 ): $h$ + $k$ = 2$n$ + 1 type seen in E4 experiment using a large single crystal glued rigidly on a sample holder are explained to be due to internal strain in the crystal. We speculate that the crystal structure of this system should be very sensitive to the application of pressure, especially around room temperature. In conclusion, the D9 neutron diffraction experiment using small single crystal free of strain shows that the high temperature crystal structure of  UPt$_2$Si$_2$ is of the expected  CaBe$_2$Ge$_2$ type. 

\acknowledgments
We would like to thank ILL for providing us with the experimental beam time within the director’s discretionary time scheme.

\end{document}